\documentclass[conference]{IEEEtran}

\usepackage[utf8]{inputenc}

\usepackage{mathtools}
\usepackage{newtxmath}
\usepackage{IEEEtrantools}
\usepackage{cite}
\usepackage{mleftright}
\mleftright

\newcommand{\mertstitle}{Waveform-Domain~NOMA:~The Future of~Multiple~Access}

\newcommand*{\matr}[1]{\mathbf{#1}}

\usepackage{dblfloatfix}
\usepackage{balance}
\usepackage[caption=false,font=footnotesize]{subfig}

\usepackage{prettyref}
\newrefformat{fig}{Fig.~\ref{#1}}
\newrefformat{conj}{Conjecture~\ref{#1}}
\newrefformat{tab}{Table~\ref{#1}}
\newrefformat{sec}{Section~\ref{#1}}
\newrefformat{subsec}{Section~\ref{#1}}
\newrefformat{subsubsec}{Section~\ref{#1}}

\usepackage{amsmath,amssymb,amsfonts}
\usepackage{algorithmic}
\usepackage[pdftex]{graphicx}
\usepackage{textcomp}
\usepackage{xcolor}

\setkeys{Gin}{width=\linewidth}

\usepackage{acronym}
\input{acros}
\usepackage{textcomp}
\usepackage{algorithmic}

\usepackage{siunitx}
\DeclareSIUnit{\belmilliwatt}{Bm}
\DeclareSIUnit{\dBm}{\deci\belmilliwatt}
\DeclareSIQualifier{\isotropic}{i}
\DeclareSIQualifier{\carrier}{c}

\makeatother

\usepackage{listings}
\def\BibTeX{{\rm B\kern-.05em{\sc i\kern-.025em b}\kern-.08em
		T\kern-.1667em\lower.7ex\hbox{E}\kern-.125emX}}
\begin{document}
	
	\title{\mertstitle}

	\author{Mehmet Mert Şahin\IEEEauthorrefmark{1}, Hüseyin
	Arslan\IEEEauthorrefmark{1}\IEEEauthorrefmark{2}\IEEEmembership{, Fellow, IEEE}\\\IEEEauthorblockA{\IEEEauthorrefmark{1}Department of Electrical Engineering, University of South Florida, Tampa, FL, 33620}\IEEEauthorblockA{\IEEEauthorrefmark{2}Department of Electrical and Electronics Engineering, Istanbul Medipol University, Istanbul, TURKEY, 34810}e-mail: mehmetmert@mail.usf.edu, arslan@usf.edu}
	
	\maketitle
	
	\begin{abstract}
	This paper introduces a novel \ac{noma} concept named waveform-domain \ac{noma}, which proposes the coexistence of different waveforms in the same \ac{re}. Regarding the demands of each \acp{ue}, appropriate waveforms are assigned intelligently and decoded at the receiver side properly. Since the performance gain of power-domain \ac{noma} compared to \ac{mumimo} degrades in the case of \ac{sic} and channel estimation errors, \ac{noma} is not considered as a work-item in 3GPP anymore. The novel waveform-domain \ac{noma} concept provides a remedy for the problems of the power-domain \ac{noma} scheme as well as flexibility in terms of reliability, latency, and energy efficiency. To mitigate \ac{sic} errors, \ac{ldpc} codes aided soft interference cancellation technique is used. As numerically evaluated, the proposed waveform-domain \ac{noma} scheme outperforms the conventional power-domain \ac{noma} in the power-balanced scenario.  
		\acresetall
	\end{abstract}	
	
	\section{Introduction}
	Over almost thirty years, tremendous efforts have been devoted to investigating new types of multiple access techniques based on the idea of serving multiple users at the same frequency, time, code, and spatial resources. Currently, these efforts are termed as \ac{noma}, which has two different techniques, including power-domain and code-domain. The main motivation behind \ac{noma} is the increased connectivity compared to \ac{oma}, which can meet the harsh requirements of the \ac{iot} \cite{NOMA_applications}. Besides studies, the concept of \ac{noma} has included in various standardization efforts. A study for the application of \ac{noma} in downlink transmission, named multi-user superposition transmission (MUST), was carried out for the 3rd Generation Partnership Project (3GPP) Release 14 \cite{3gpp.36.859}. A study for the application of \ac{noma} for uplink transmission has been recently carried out for 3GPP Release 16, where different implementations of \ac{noma} have been studied \cite{3gpp.38.812}. However, since power-domain \ac{noma} has performance degradation in some cases mentioned throughout the paper, it is not considered as a work-item in Release 17. 
	
	Power-domain \ac{noma} can yield much better spectral efficiency compared to \ac{oma} unless channel gains of users are not similar \cite{NOMA_myths}. To date, the literature on power-domain \ac{noma} has based its work on perfect \ac{sic} process \cite{pd_NOMA}, i.e., there is no residual error during the detection, reconstruction, and subtraction of the decoded user's waveform from the superimposed received signal. However, this assumption is infeasible because the signal of the firstly decoded user should be estimated perfectly at the receiver \cite{JIC_MCM_NOMA}. Additionally, although the impact of channel estimation accuracy on the performance of \ac{bler} is negligible for downlink \ac{noma} \cite{recDesign_DL_NOMA}, it should be considered for uplink \ac{noma} regarding the reconstruction of firstly decoded user’s waveform. On the other hand, the transmit power of users is arranged in a way that the users' power received at the \ac{bs} is significantly different in order to enhance the overall system throughput in power-domain \ac{noma} \cite{PA_ULandDL_NOMA}. Since users transmitting at a similar power level may also be grouped in the case of high connectivity, various researches have been conducted to find new \ac{noma} schemes that operate in power-balanced scenarios \cite{ncma_noma}. 
	
	A novel uplink \ac{noma} scheme with two users based on \ac{ofdm} and \ac{ofdmim} is proposed in \cite{OFDM-IM_NOMA} where inherent power imbalance of \ac{ofdmim} leads to better throughput compared to conventional \ac{ofdm} and \ac{ofdm} \ac{noma} in the case of power-balanced scenario. However, the throughput performance of this scheme is not evaluated in the case of various power differences between the user's received signal. Besides, \ac{noma} with codeword-level \ac{sic} is studied in \cite{sdr_NOMA} with turbo decoding to improve the reliability of the firstly decoded signal. However, in this study, the demodulated bits are reencoded and then subtracted from the superimposed received signal to perform demodulation for the second user. The catastrophic errors that would occur in the case of reencoding the falsely decoded bits are not considered. 
		
	This paper proposes a novel \ac{noma} scheme called waveform-domain \ac{noma} with transceiver design utilizing \ac{ldpc} codes aided soft interference cancellation to improve \ac{bler} performance through all received power level variations of users. The \ac{llr} calculations are evaluated depending on the waveform type that is decoded first. The reconstruction of the firstly decoded user's waveform is investigated with two different techniques in terms of \ac{evm} as well as considering channel estimation errors. It is shown that the proposed waveform-domain \ac{noma} scheme outperforms the conventional power-domain \ac{noma} scheme in terms of \ac{bler} performance in the power-balanced scenarios. Moreover, waveform-domain \ac{noma} provides flexibility among users regarding their demands.   
	
	\section{System Model}
	For brevity, consider two users uplink \ac{noma} scenario where both users transmit to a single \ac{bs} over $N$ subcarriers in the presence of frequency selective channels, including \ac{awgn}. Although the uplink scenario is considered, the waveform-domain \ac{noma} concept is also applicable for downlink transmission.  
	\begin{figure}
		\centerline{\includegraphics{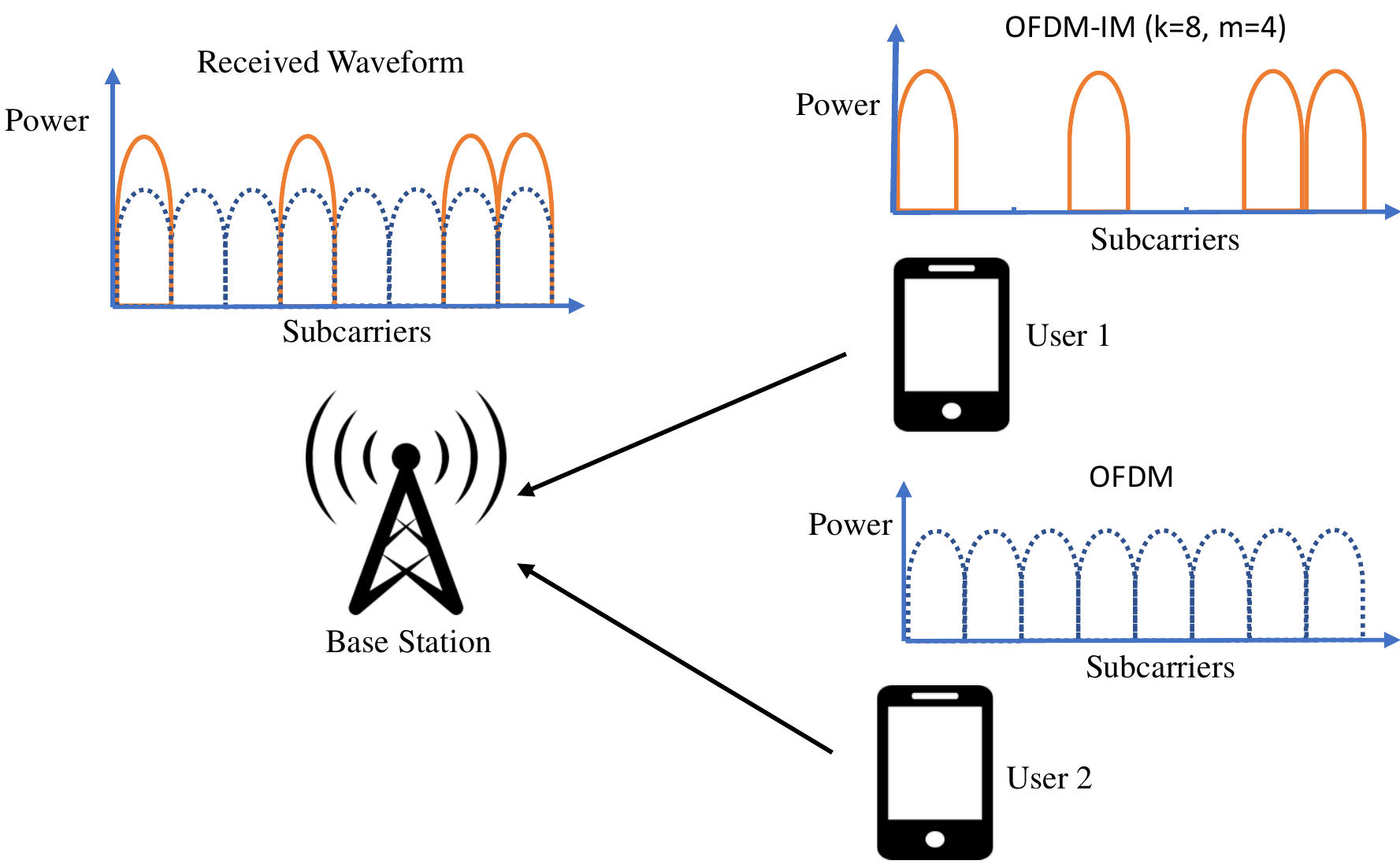}}
		\caption{Two users uplink \ac{noma} scheme with \ac{ofdmim}+\ac{ofdm}.}
		\label{fig:twoUserScheme} \hfill
	\end{figure}
	
	\subsection{Conventional power-domain \ac{noma} with \ac{ofdm}+\ac{ofdm} \label{sec:ofdm_ofdm}} 	
	Firstly, both \acp{ue} encode their messages by \ac{ldpc} codes and modulate via \ac{qam}, where data symbols of users are drawn from a complex symbol alphabet $\mathbb{S}$. Then, these symbols are \ac{ofdm} modulated and transmitted to be received by the \ac{bs} over the same \ac{re}. Moreover, $p_1$ and $p_2$ denote the signal power of user 1 and user 2 for each subcarrier, respectively. In the \ac{ofdm} with total $N$ subcarriers, the total powers of user 1 and user 2 become $P_{1} = Np_{1}$ and $P_{2} = Np_{2}$, respectively. After the process of \ac{fft} and removal of cyclic prefix, the baseband received signal at the $n$th subcarrier is expressed as follows:	
	\begin{equation}
	r_n = \sqrt{p_1} h_{1,n} u_{1,n} + \sqrt{p_2} h_{2,n} u_{2,n} + w_n,
	\end{equation} 
	where $h_{1,n}$, $h_{2,n}$, $u_{1,n}$, and $u_{2,n}$ are the channel gains and data symbols of users 1, and 2, respectively. Also, $w_n\sim\mathcal{CN}\left(0,\sigma^2\right)$ denotes the \ac{awgn} at the $n$th subcarrier.
	
	Assuming that the signal of user 1 is decoded first, the capacity of user 1 ($R_1$) in conventional power-domain \ac{noma} is given by	
	\begin{equation} \label{eq:dataRateOFDMFirst}
	R_1 = \sum_{n=1}^N\log_2\left(1+\frac{p_{1}h_{1,n}}{\sigma^2+p_{2}h_{2,n}}\right) 
	\text{bit/s/Hz.}
	\end{equation}	  
	Assuming perfect \ac{sic}, which is infeasible, the capacity of user 2 ($R_2$) is calculated as follows: 	
	\begin{equation} \label{eq:dataRateOFDMSecond}
	R_2 = \sum_{n=1}^N\log_2\left(1+\frac{p_{2}h_{2,n}}{\sigma^2}\right) 
	\text{bit/s/Hz.}
	\end{equation}
	In the case of \ac{mlmud} without \ac{sic}, the decoding order does not have any effect on the sum-rate; therefore, any arbitrary decoding order can be assumed to be performed \cite{uplink_NOMA}. On the other hand, when \ac{ml}-\ac{mud} with \ac{sic} is used, the stronger user should be decoded first. 	

	\subsection{\ac{noma} with \ac{ofdmim}+\ac{ofdm} \label{sec:ofdm_ofdm-im}}	
\begin{figure}
		\subfloat[]{
			\includegraphics{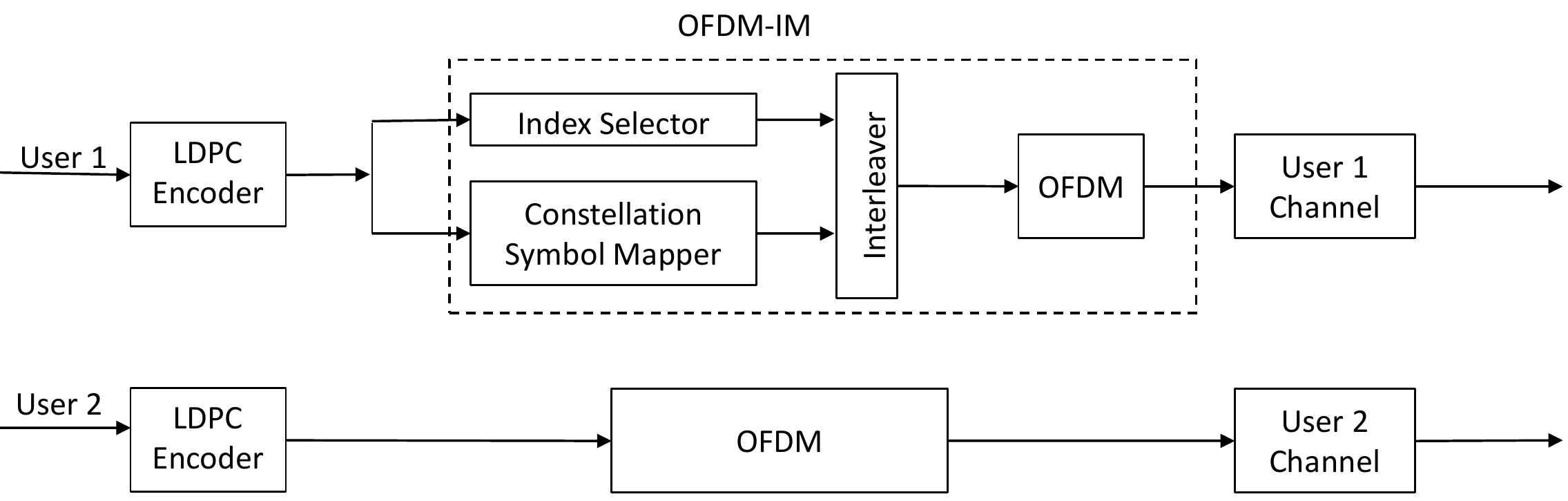}
			\label{fig:gen_frame1}}\hfil
		\subfloat[]{
			\includegraphics{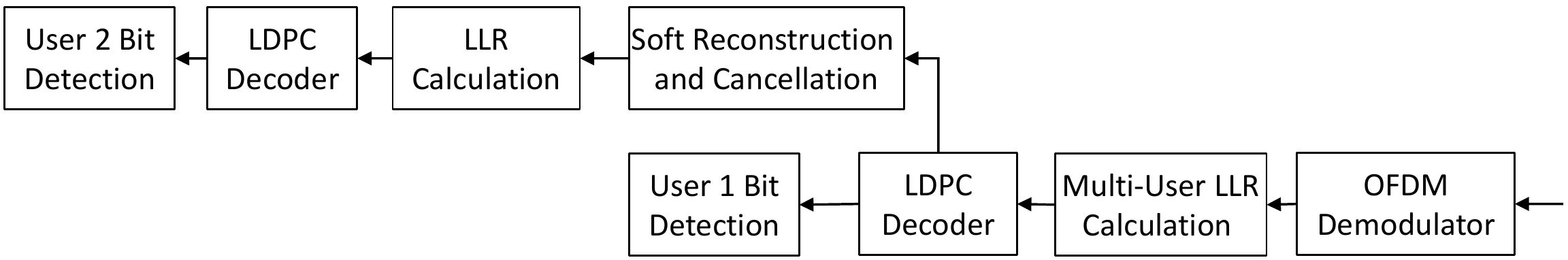}
			\label{fig:gen_frame2}}\hfil
		\caption{Proposed transmission and reception scheme (a) Coding and modulating of user 1 (\ac{ofdmim}) and user 2 (\ac{ofdm}) signals, (b) Demodulating and decoding of the superimposed received signal with \ac{ldpc} codes aided soft reconstruction and cancellation.}
		\label{fig:symstr}
	\end{figure}

	As \prettyref{fig:twoUserScheme} depicts, user 1 has utilized \ac{ofdmim} waveform, whereas user 2 sends its signal via \ac{ofdm} waveform over $N$ subcarriers. In the \ac{ofdmim} scheme \cite{ofdm_IM}, the total $Q=Q_1+Q_2$ bits are transmitted as follows: Firstly, $N$ subcarriers are split into $g$ subblocks consisting of $k$ subcarriers. The $Q_1$ bits are used to determine the indices of $m$ active subcarriers where the total number of active subcarrier positions is denoted as $c$. In each subblock $\beta$, only $m$ out of $k$ subcarriers have activated.  
	Activated subcarriers are used to map $Q_2$ bits on to $M$-ary signal constellation symbols selected from the complex set $\mathbb{S}$. The information of user 1 carried in the subblock $\beta$ is given by $\matr{u_{1,\beta}} = \left[u_{1,\beta}^{(1)} \ldots u_{1,\beta}^{(Q)}\right]$. As seen in \prettyref{fig:gen_frame1}, the interleaved grouping is performed to increase the achievable rate of \ac{ofdmim} \cite{rate_OFDM-IM}. In subblock $\beta$, the vector of modulated symbols of user 2 carried with \ac{ofdm} waveform is denoted by $\matr{u_{2,\beta}} = \left[u_{2,\beta}^{(1)} \ldots u_{2,\beta}^{(k)}\right]$.	
	
	After \ac{fft} and cyclic prefix removal, the superimposed received signal at the $n$th subcarrier becomes   
	\begin{equation} \label{eq:sysModel}
	r_n = \sqrt{\frac{k P_1}{m N}} h_{1,n} u_{1,n} + \sqrt{p_2} h_{2,n} u_{2,n} + w_k, 
	\end{equation}
	where $ u_{1,n} \in \mathbb{S^\prime}=\{0,\mathbb{S}\}$. Moreover, denote $\matr{r}_{\beta} \in \mathbb{C}^{1\times k}$ as the received signal at the $\beta$th subgroup. \prettyref{fig:gen_frame2} depicts the reception process of the proposed \ac{noma} scheme by decoding the \ac{ofdmim} waveform first. However, the decoding may not always start with the \ac{ofdmim} waveform. It depends on both power, subcarrier allocation, and modulation order.
	
	\section{\Ac{llr} Calculations}
	This section includes the \ac{llr} calculations of each user for two different \ac{noma} schemes. Calculated \acp{llr} are sent to the \ac{ldpc} decoder as input. For the sake of fair comparison, we have used the log-sum approximation technique \cite{BICM} to calculate approximate \acp{llr} of two different \ac{noma} schemes.
		 	
	\subsection{\ac{llr} calculations for \ac{noma} with \ac{ofdm}+\ac{ofdm}}	
	With \ac{ml}-\ac{mud} algorithm, the \ac{llr} of the bit $i$ of user 1 at the $n$th subcarrier, $\Lambda_{n^{(i)}}^{u_1}$, is calculated as
	\begin{IEEEeqnarray}{rCl} \label{LLR_OFDM_OFDM}
	\Lambda_{n^{(i)}}^{u_1}&=&\log \left( \frac{f(r_n | u_{1,n}^{(i)}=0)}{f(r_n | u_{1,n}^{(i)}=1)} \right) \nonumber \\
	& \approx & \min_{u_{1,n} \colon u_{1,n}^{(i)}\in \mathbb{S}_{1}^i,\ u_{2,n}\in \mathbb{S}} \frac{\|r_n-h_{1,n} u_{1,n}-h_{2,n} u_{2,n}  \|^2}{\sigma^2}  \nonumber \\ 
	&& - \min_{u_{1,n} \colon u_{1,n}^{(i)}\in \mathbb{S}_{0}^i, \ u_{2,n}\in \mathbb{S}} \frac{\| r_n-h_{1,n} u_{1,n}-h_{2,n} u_{2,n} \|^2}{\sigma^2},\IEEEeqnarraynumspace 
	\end{IEEEeqnarray} 
	where $\mathbb{S}_{b}^i \subset \mathbb{S}$ denotes the set of all symbols $\alpha \in \mathbb{S}$ whose label has $b \in \{0,1\}$ in bit position $i$. The complexity of this \ac{llr} calculation, in terms of complex multiplications, becomes $\sim \mathcal{O}\left(\left|\mathbb{S}\right|^{2}\right)$.
	After \ac{ldpc} decoder fed with \acp{llr}, the symbols of user 1 is reconstructed and subtracted from the superimposed signal with inevitable \ac{sic} error. The \acp{llr} of user 2 are calculated with the remaining signal and sent to the \ac{ldpc} decoder in order to obtain bit decisions of user 2. 
	
	\subsection{\ac{llr} calculations for \ac{noma} with \ac{ofdmim}+\ac{ofdm}}
	The \ac{llr} calculations for users' bits in the \ac{ofdmim}+\ac{ofdm} \ac{noma} scheme depend on which waveform is decided to be decoded first. As it is shown via numerical results in \prettyref{sec:sim_section}, the total power level is not the unique limitation to decide which waveform should be decoded first. By decoding the \ac{ofdmim} waveform first, the \ac{llr} of the bit $i$ of user 1 at the $\beta$th subgroup, $\Lambda_{\beta^{(i)}}^{u_1}$ is  
	\begin{IEEEeqnarray}{rCl} 
		\IEEEeqnarraymulticol{3}{l}{\Lambda_{\beta^{(i)}}^{u_1}}  \nonumber \\
		&=&\log \left( \frac{f(\matr{r}_{\beta} | u_{1,\beta}^{(i)}=0)}{f(\matr{r}_{\beta} | u_{1,\beta}^{(i)}=1)} \right)  \nonumber\\
		&\approx& \min_{\matr{u}_{1,\beta} \colon u_{1,\beta}^{(i)}=1,\ \matr{u}_{2,\beta} \in \{S\}^k} \frac{\| \matr{r}_\beta-\matr{h}_{1,\beta}\odot \matr{u_{1,\beta}}-\matr{h}_{2,\beta}\odot\matr{u}_{2,\beta} \|^2}{\sigma^2}  \nonumber \\ &&
		\IEEEeqnarraymulticol{1}{r}{
		 -\>\min_{\matr{u}_{1,\beta} \colon u_{1,\beta}^{(i)}=0,\matr{u}_{2,\beta} \in \{S\}^k} \frac{\| \matr{r}_\beta-\matr{h}_{1,\beta}\odot\matr{u_{1,\beta}}-\matr{h}_{2,\beta}\odot \matr{u}_{2,\beta}\|^2}{\sigma^2}\IEEEeqnarraynumspace} 
	\end{IEEEeqnarray} 	
	where $\matr{h}_{1,\beta} \in \mathbb{C}^{1\times k}$ and $\matr{h}_{2,\beta} \in \mathbb{C}^{1\times k}$ denote the \ac{csi} of users 1 and 2 through $\beta$th subgroup, respectively, and $\odot$ denotes Hadamard multiplication.
	When the \ac{ofdmim} waveform is decoded first, the complexity of \ac{llr} calculation, in terms of complex multiplications, becomes $\sim\mathcal{O}\left(c\left|\mathbb{S}\right|^{m} \left|\mathbb{S}\right|^{k}\right)$. On the other hand, starting the decoding process with the \ac{ofdm} waveform, the \ac{llr} of the bit $i$ of user 2 at the $n$th subcarrier, $\Lambda_{n^{(i)}}^{u_2}$, becomes	
	\begin{IEEEeqnarray}{rCl} \label{LLR_OFDM_IM}
		\Lambda_{n^{(i)}}^{u_2} & = &\log \left( \frac{f(r_n | u_{2,n}^{(i)}=0)}{f(r_n | u_{2,n}^{(i)}=1)} \right) \nonumber \\
		& \approx & \min_{u_{2,n} \colon u_{2,n}^{(i)}\in \mathbb{S}_{1}^i,\ u_{1,n}\in \mathbb{S}^{\prime}} \frac{\|r_n-h_{2,n} u_{2,n}-h_{1,n} u_{1,n}  \|^2}{\sigma^2} \nonumber \\ && -  \min_{u_{2,n} \colon u_{2,n}^{(i)}\in \mathbb{S}_{0}^i, \ u_{1,n}\in \mathbb{S}^{\prime}} \frac{\| r_n-h_{2,n} u_{2,n}-h_{1,n} u_{1,n} \|^2}{\sigma^2}. \IEEEeqnarraynumspace 
	\end{IEEEeqnarray} 
	By decoding the \ac{ofdm} waveform first, the complexity of \ac{llr} calculation, in terms of complex multiplications, becomes $\sim\mathcal{O}\left(\left|\mathbb{S^{\prime}}\right|\left|\mathbb{S}\right|\right)$. The waveform, whichever is decoded first, is reconstructed and subtracted from the aggregate received signal before the next user's signal is decoded. 

	\section{Waveform Domain \ac{noma}}
	The waveform consisting of symbol, pulse shape, and lattice is the physical shape of the signal carrying modulated information \cite{MulticarrierComm_Survey}. The novel waveform-domain \ac{noma} concept proposes the combination of different waveforms along with the non-orthogonal resources to introduce flexibility, separability, and detectability.
	
	\subsection{Waveform Selection}	
	Some intelligent techniques for overlapping of users with different waveforms are studied in the prior works \cite{Sari_CDMAandOFDM,MBC_OFDMandSCFDMA,OTFS_NOMA}. In \cite{Sari_CDMAandOFDM}, the separability aspect on the overlapping of two different waveforms, which are \ac{ofdm} and \ac{cdma}, is investigated with an iterative receiver design that is computationally complex. Similarly, the coexistence of \ac{ofdm} and \ac{sc-fdma} is studied in \cite{MBC_OFDMandSCFDMA}. It is shown that the proposed \ac{mud} approach utilizing iterative likelihood testing and \ac{sinr} based processing outperforms conventional \ac{sic}. Moreover, the \ac{otfs} waveform is used for the high mobility user, whereas the signal of the low mobility user is transmitted via the \ac{ofdm} waveform in \cite{OTFS_NOMA}. This \ac{noma} concept provides flexibility among users according to their mobility profiles. In our study, due to some superiority of \ac{ofdmim} over \ac{ofdm} such as ergodic achievable rate, \ac{papr} reduction, and robustness to \ac{ici} \cite{IM_techniques}, \ac{ofdmim} is chosen as a candidate for waveform-domain \ac{noma}.    
	
	\subsection{Reconstruction of the Waveform}
	In the uplink \ac{noma}, errors in channel estimation and information detection propagate through the users coming in successive decoding order. To mitigate the performance degradation, we have used \ac{ldpc} codes aided soft-interference cancellation. The soft reconstruction scheme is compared with the reencoding scheme proposed in \cite{sdr_NOMA} in terms of \ac{evm} performance. 		
	\ac{evm} is calculated as the ratio of \ac{rms} power of the error vector to the \ac{rms} power of the reference symbol overall \ac{ofdm} subcarriers. For user 1, \ac{evm} is calculated as follows:	
	\begin{IEEEeqnarray}{rCl} \label{evm}
		\text{EVM(dB)} & = & 10\log_{10}\bigg(\sqrt\frac{\sum_{n=1}^N \| {\hat{u}_{1,n}}-{u_{1,n}}\|^2}{\sum_{n=1}^N \|{u_{1,n}}\|^2}\bigg), \nonumber
	\end{IEEEeqnarray} 	
	where $\hat{u}_{1,n}$ corresponds to the reconstructed symbol of user 1 at the $n$th subcarrier. Two different techniques are used to reconstruct the decoded waveform to be stripped away from the superimposed received signal:  
	\begin{enumerate}
		\item If the signal of user 1 is demodulated first, the decoded bits of user 1 are reencoded and remodulated to reconstruct the waveform of user 1. Then it is subtracted from the superimposed signal to decode the signal of user 2. 
		\item For the reconstruction of the firstly decoded signal, we have utilized the Gallager sum-product algorithm \cite{channelCodes_classicalANDmodern} that provides the total \ac{llr} of each variable node at the stopping stage. The total \ac{llr} shows how reliable the symbol consisting of the related bit can be reconstructed. Soft decision reconstruction is done at the end of the iterative process of the sum-product algorithm over the variable node $v_i$ as follows: 		
		\begin{equation}
		\zeta = \tanh(L_i),
		\end{equation}			
		where $\zeta$ denotes the scale factor for the symbol consisting of the related bit $i$ and $L_i$ is the sum of the extrinsic \acp{llr} from connected check nodes and initial \ac{llr} of the bit $i$. For example, in BPSK modulation, the reconstructed signal of user 2 for the $i$th bit on the $n$th subcarrier becomes $\hat{u}_{2,n}=-\sqrt{p_2}\tanh(L_i)$. Subsequently, the reconstructed signal is canceled from the aggregate received signal to decode the signal of another user.		
	\end{enumerate}		
	
\begin{figure}
		\centerline{\includegraphics{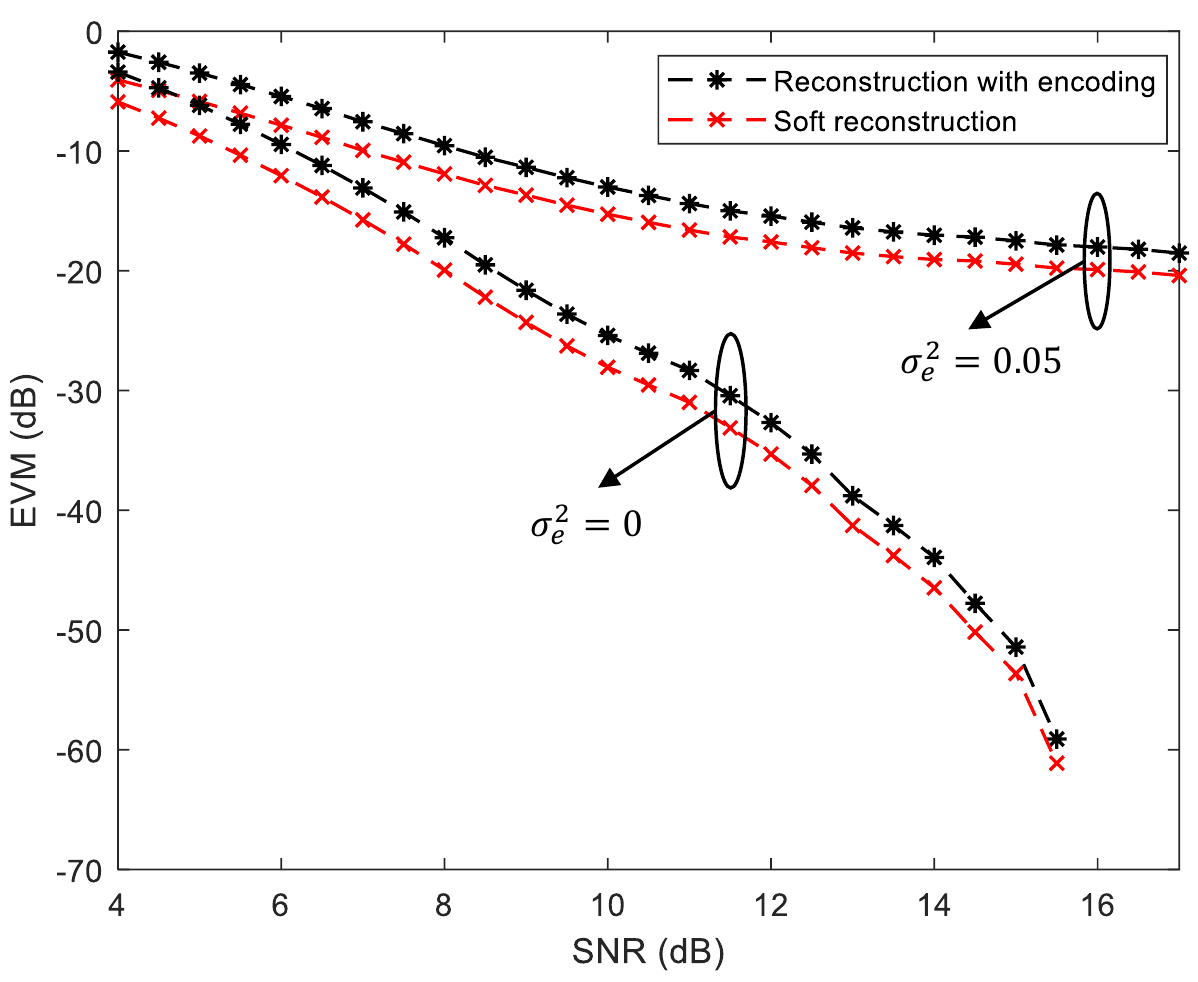}}
		\caption{Comparison of reconstruction schemes for different \acp{mse} ($\sigma_e^2$) in channel estimation.}
		\label{fig:reconsSchemes} \hfill
	\end{figure}

	Comparison of different reconstruction schemes is evaluated through Monte Carlo simulations over the Frequency Selective Channel with 10 taps. \Ac{pdp} of the channel is considered as uniform. Moreover, \ac{mse} is defined as the expected value of the normalized difference between the channel response $h$ and the channel estimation $\tilde{h}$, $\sigma_e^2 = \frac{\mathbb{E}[\mid\tilde{h}-h\mid^2]}{\mathbb{E}[\mid h \mid^2]}$, where $\mathbb{E[\cdot]}$ denotes the expected value. In \prettyref{fig:reconsSchemes}, \ac{evm} results are provided for different \acp{mse} in channel estimation. It portrays the superiority of the proposed reconstruction scheme with extrinsic soft \acp{llr} information obtained from the \ac{ldpc} decoder. It can also be seen that, in the uplink \ac{noma}, accuracy in channel estimation has a high impact on the reconstruction of the firstly decoded user's waveform.  
	\section{Simulation Results and Discussion \label{sec:sim_section}}
	
	The proposed technique is evaluated numerically through Monte Carlo simulations. As modulation order, QPSK signaling is used for both two \ac{noma} schemes, where the equal data rate is satisfied with three active subcarriers in the group of four subcarriers ($m=3$, $k=4$) for the user utilizing the \ac{ofdmim} waveform. For \ac{ofdm}+\ac{ofdm} \ac{noma}, the user with high received power is decoded first, then reconstructed, and canceled from the superimposed signal. On the other hand, for \ac{ofdmim}+\ac{ofdm} \ac{noma}, decoding order is determined according to waveform type. Firstly decoded waveform is shown as bold for all given plots. 
	\begin{figure}
		\subfloat[]{
			\includegraphics{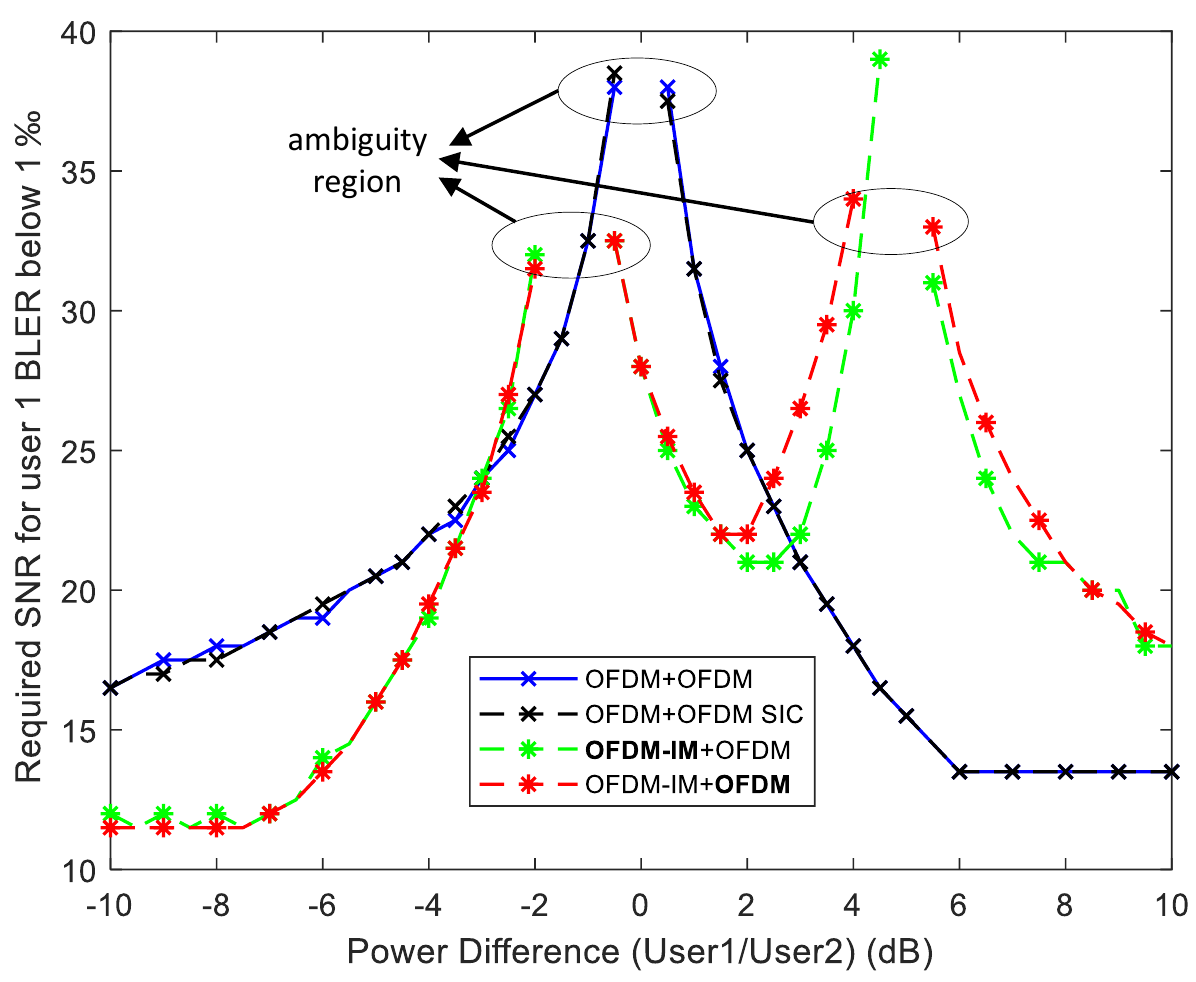}
			\label{fig:firstUserRAW_AWGN}}\hfil
		\subfloat[]{
			\includegraphics{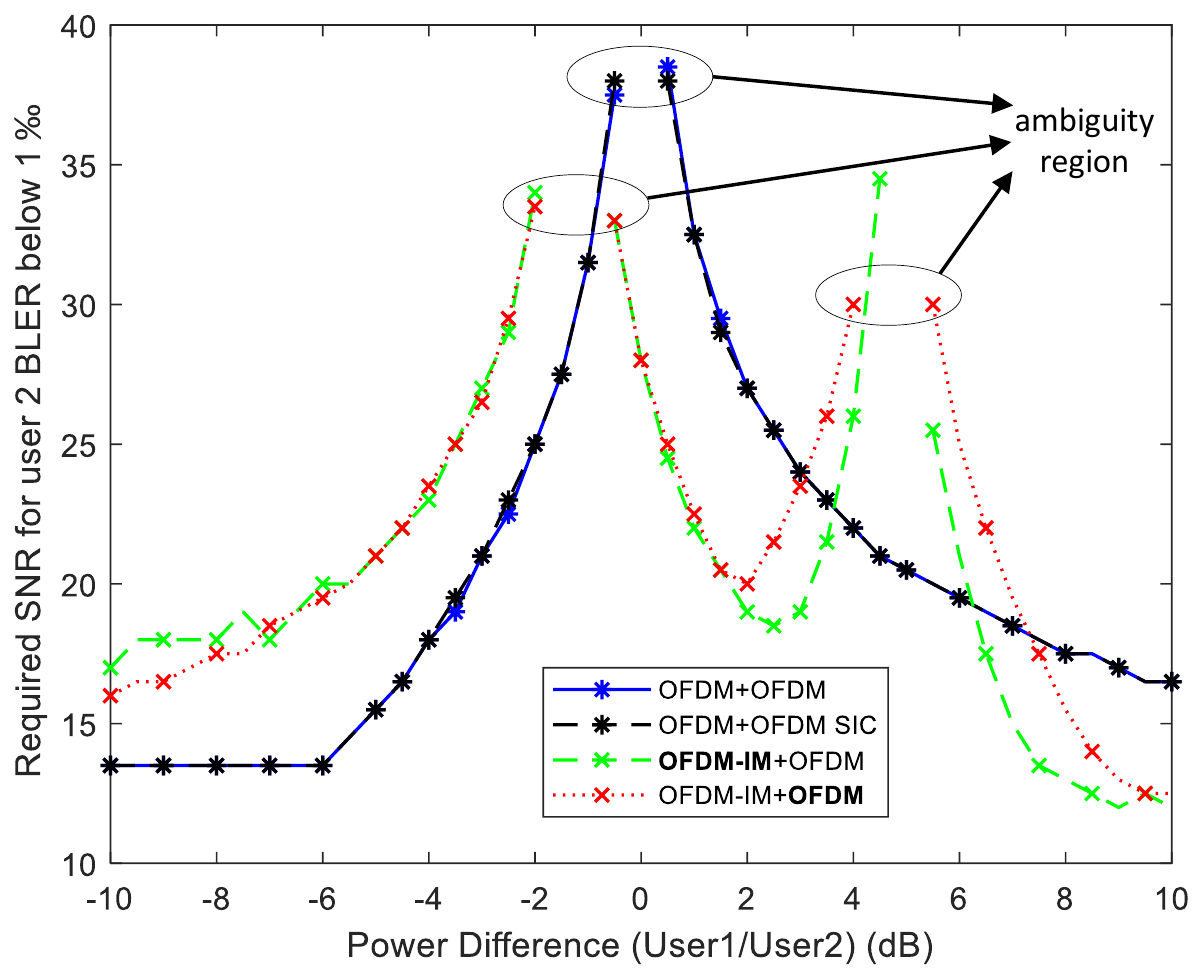} \hfil
			\label{fig:secondUserRAW_AWGN}}
		\vspace{-0.2cm}
		\caption{Comparison of uncoded \ac{noma} schemes in \ac{awgn} (a) \ac{bler} of user 1 (b) \ac{bler} of user 2.} 
		\label{fig:rawAWGN} 
		\vspace{-0.25cm}
	\end{figure}	
	\begin{figure}
		\subfloat[]{
			\includegraphics{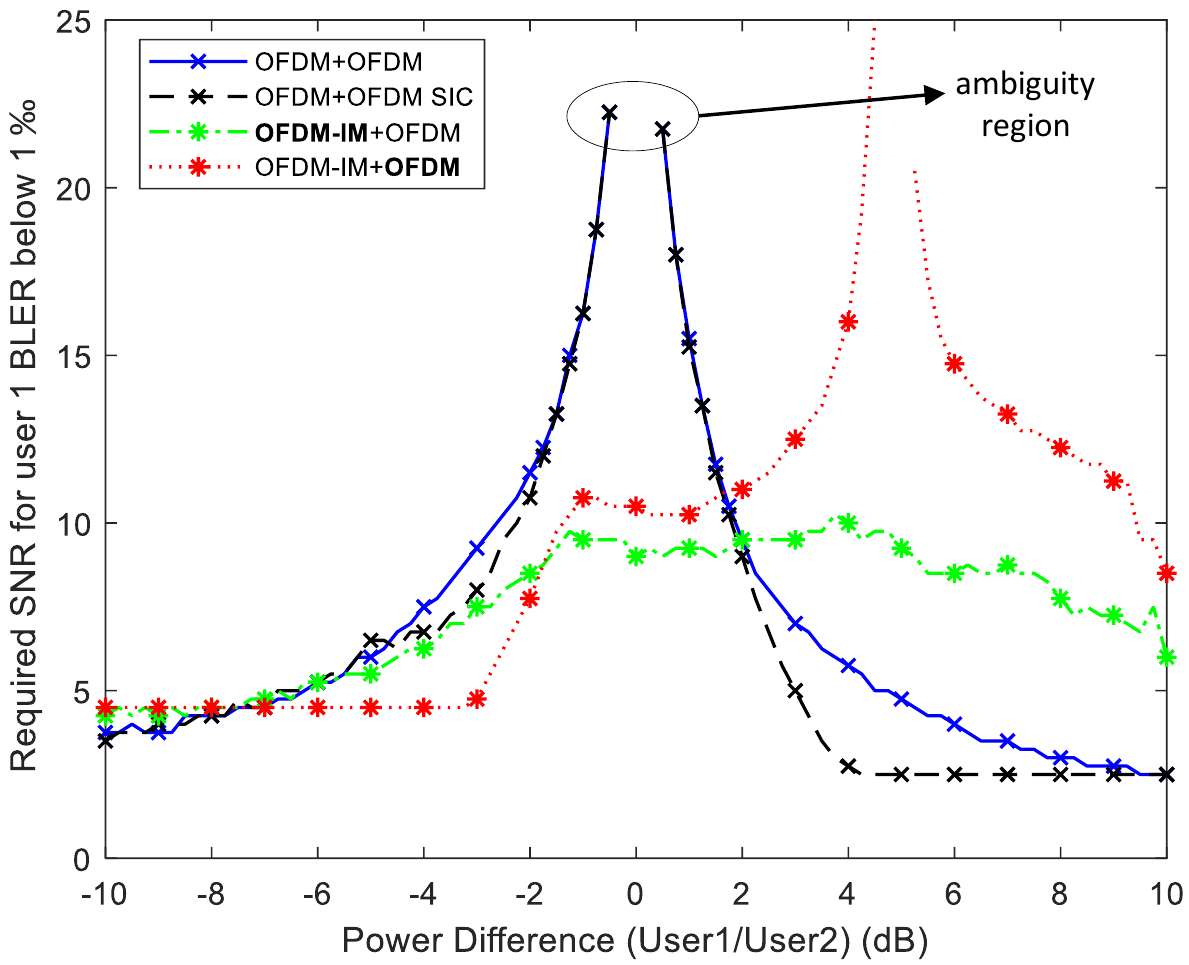}
			\label{fig:firstUserCODED_AWGN}}\hfil
		\subfloat[]{
			\includegraphics{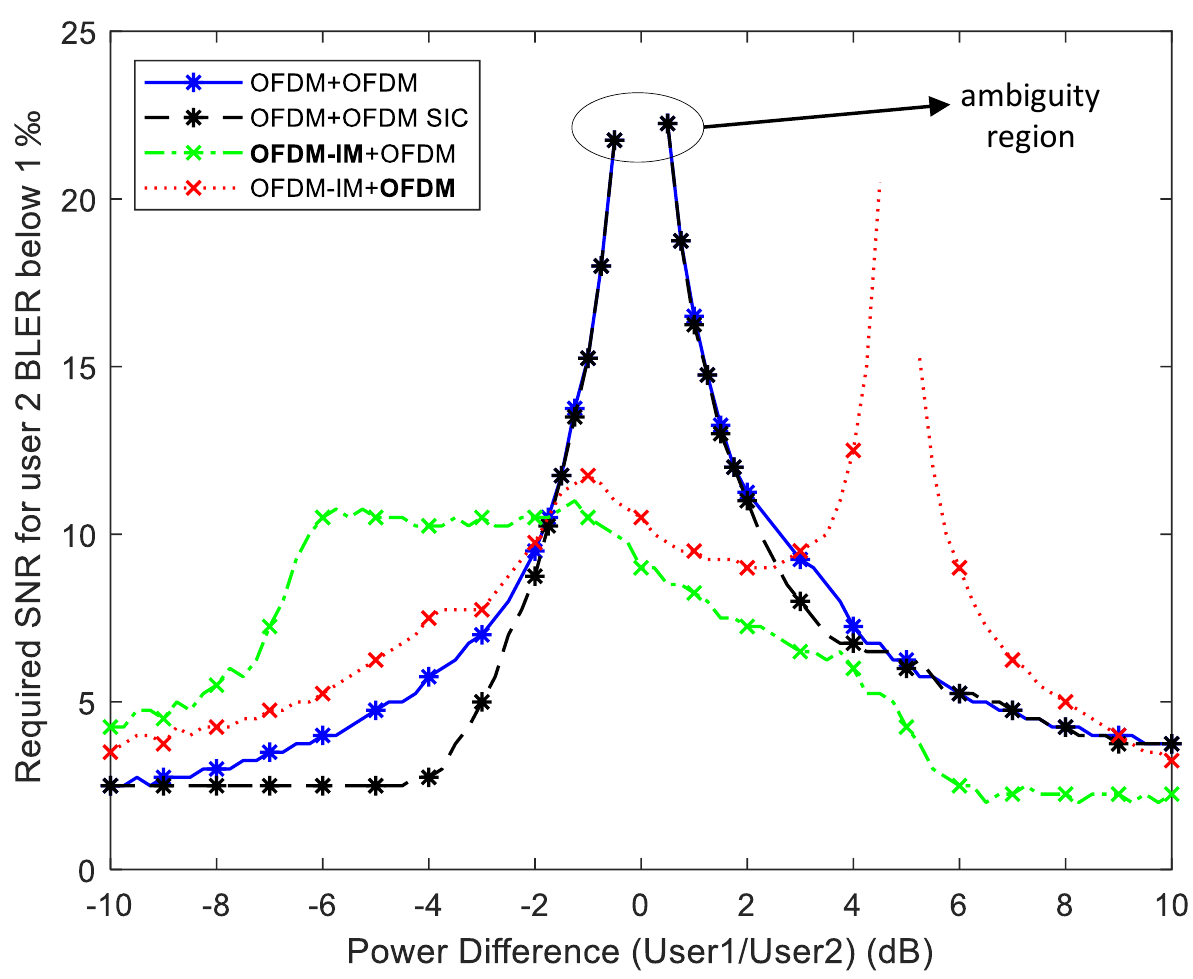} \hfil
			\label{fig:secondUserCODED_AWGN}}
		\vspace{-0.2cm}
		\caption{Comparison of coded conventional power-domain and proposed waveform-domain \ac{noma} schemes in \ac{awgn} (a) \ac{bler} of user 1 (b) \ac{bler} of user 2.} 
		\label{fig:codedAWGN} 
		\vspace{-0.25cm}
	\end{figure}  

	\prettyref{fig:firstUserRAW_AWGN} and \prettyref{fig:secondUserRAW_AWGN} demonstrate the uncoded performance of both \ac{noma} schemes over the \ac{awgn} channel for user 1 and user 2, respectively. The vertical axis denotes the required \ac{snr} for a user to achieve the target \ac{bler} of $1\text{\textperthousand}$, whereas the horizontal axis denotes the power difference in terms of \si{dB} between two different users. \ac{ofdm}+\ac{ofdm} \ac{noma} performance degrades significantly when users' power is close to each other. As reference \cite{OFDM-IM_NOMA} shows that, \ac{ofdmim}+\ac{ofdm} \ac{noma} is superior in terms of \ac{bler} at the region, where power difference between users is very close to \SI{0}{\deci\bel}. This superiority comes from the inherent power difference of the \ac{ofdmim} structure. However, as power imbalance between the users nears \SIlist{-2;5}{\deci\bel}, the performance degrades significantly because power coefficients $\sqrt{\frac{kP_1}{mN}}$ and $\sqrt{p_2}$ in \prettyref{eq:sysModel} equate the aggregate to the decision boundary. These regions are called as ambiguity region where user's messages are not decoded even with high \ac{snr}.   	

	\prettyref{fig:firstUserCODED_AWGN}, and \prettyref{fig:secondUserCODED_AWGN} compare the \ac{ldpc} coded conventional power-domain and the proposed waveform-domain \ac{noma} with 0.5 code rate for users 1 and 2, respectively. The block length is chosen as 256. For waveform-domain \ac{noma}, the decoding order should be chosen properly to enhance the performance gain. As opposed to conventional power-domain \ac{noma}, the superior region of the user 1 and user 2 is roughly below \SI{2}{\deci\bel} and above \SI{-2}{\deci\bel}, respectively. Using forward error correction with soft reconstruction and cancellation removes deep performance losses in the range of certain power differences for \ac{ofdmim}+\ac{ofdm} \ac{noma} scheme. However, conventional power-domain \ac{noma} still has a region where the performance degrades significantly. The proposed waveform-domain \ac{noma} scheme is superior at the region where the power of users is close to each other without having significant performance losses as the power difference between users increases. 
	\begin{figure}
		\centerline{\includegraphics{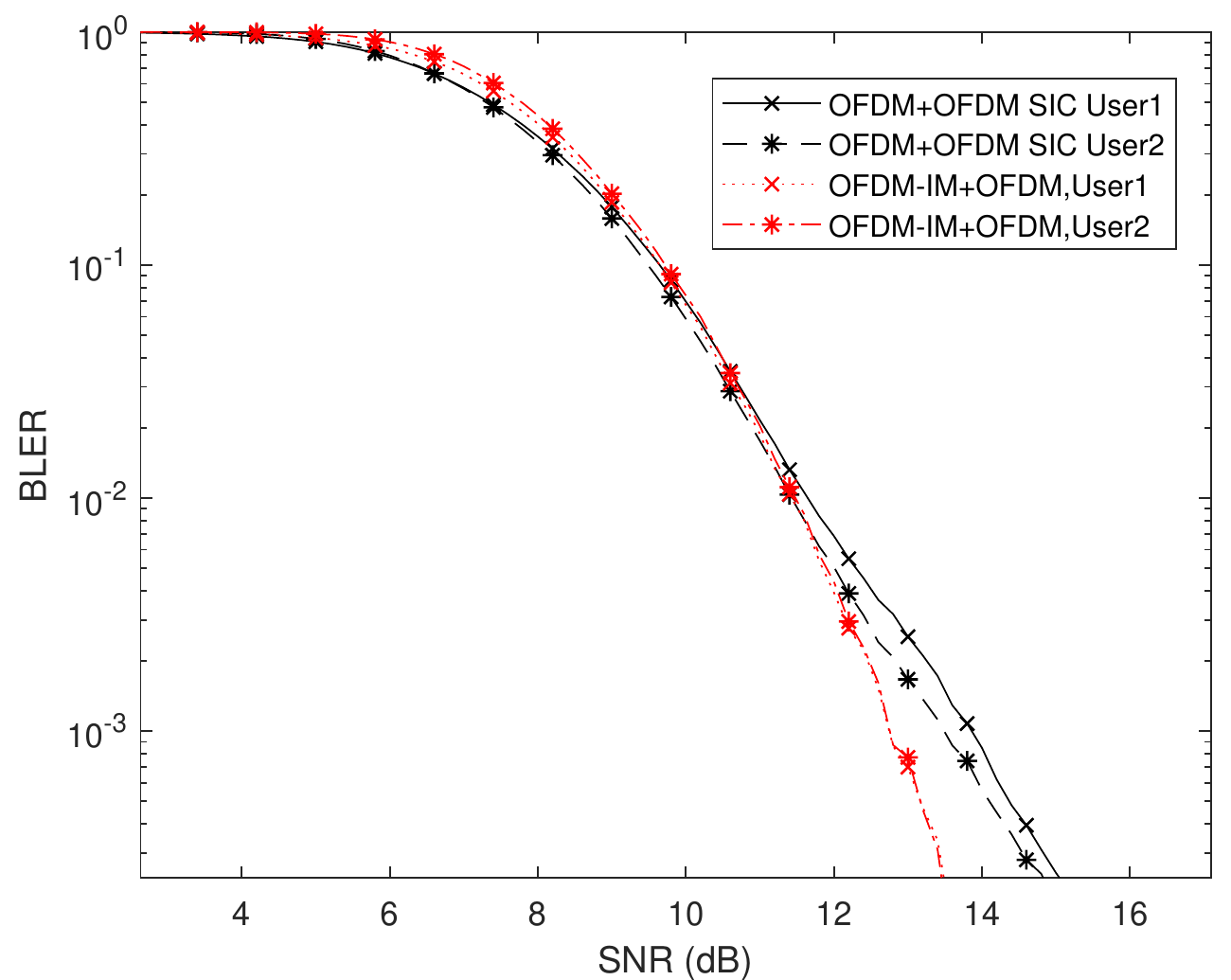}}
		\vspace{-0.2cm}
		\caption{Comparison of conventional and proposed \ac{noma} schemes in frequency selective channel with tap number is 10.}
		\label{fig:Rayleigh_BER_CODED} \hfil
		\vspace{-0.25cm}
	\end{figure}
	
	\prettyref{fig:Rayleigh_BER_CODED} depicts the \ac{bler} performance of the conventional power-domain and the proposed waveform-domain \ac{noma} schemes through frequency selective channel with 10 taps where users modulate their signal with equal power. Throughout the simulation, it is assumed that channel knowledge is present at the receiver. Since the channel is frequency selective, the received signal has different power imbalances over each subcarrier; in other words, power equality does not hold any more at the receiver side. Even in this case, better performance is obtained with the proposed waveform-domain \ac{noma} scheme. Nearly \SI{1}{\deci\bel} gain  is achieved for both users at the target \ac{bler} of $1\text{\textperthousand}$.
	\section{Conclusions}
	In this paper, the novel waveform-domain \ac{noma} concept is presented and compared with the conventional power-domain \ac{noma} scheme. The numerical results demonstrate that the proposed waveform-domain \ac{noma} scheme is capable of overcoming the problems of power-domain \ac{noma} in power-balanced scenarios. Moreover, waveform-domain \ac{noma} gives flexibility to the users regarding their demands such as reliability, energy efficiency, spectral efficiency, and latency. A promising future research direction is to investigate the optimal waveforms that can be paired in waveform-domain \ac{noma}. Also, the optimal coding schemes that are convenient for the chosen waveforms may be studied. The promising results of the waveform-domain \ac{noma} concept may potentially spur the interest of the wireless industry, and academia; and pave the way for being possible multiple access scheme of 6G and beyond. 
	
	\section{Acknowledgement}
	This work is supported by the National Science Foundation under Grant ECCS-1609581. 
	\bibliographystyle{IEEEtran}
	\bibliography{mert,3gpp_38-series}
	
\end{document}